\documentclass[eqsecnum,aps,prb,twocolumn]{revtex4}

\usepackage{graphicx}%
\usepackage{dcolumn}
\usepackage{amsmath}

\makeatletter
\def\btt#1{\texttt{\@backslashchar#1}}%
\DeclareRobustCommand\bblash{\btt{\@backslashchar}}%
\makeatother

\begin{document}

\title[]{Fermi surface of the heavy fermion superconductor PrOs$_{4}$Sb$_{12}$}

\author{H.~Sugawara,~$^{1}$ S.~Osaki,~$^{1}$ S.~R.~Saha,~$^{1}$ Y.~Aoki,~$^{1}$ H.~Sato,~$^{1}$ Y.~Inada,~$^{2}$ H.~Shishido,~$^{2}$ R.~Settai,~$^{2}$ Y.~\={O}nuki,~$^{2,3}$ H.~Harima,~$^{4}$ and K.~Oikawa~$^{5}$}

\affiliation{
$^1$Graduate School of Science, Tokyo Metropolitan University, Minami-Ohsawa, Hachioji, Tokyo 192-0397, Japan\\
$^2$Graduate School of Science, Osaka University, Toyonaka, Osaka, 560-0043, Japan\\
$^3$Advanced Science Research Center, Japan Atomic Energy Research Institute, Tokai, Ibaraki 319-1195, Japan\\
$^4$The Institute of Scientific and Industrial Research, Osaka University, Ibaraki, Osaka 567-0047, Japan\\
$^5$Neutron Science Laboratory, Institute of Materials Structure Science, High Energy Accelerator Research Organization, Tsukuba, 305-0801, Japan
}

\date{\today}

\begin{abstract}
~~~We have investigated the de Haas-van Alphen effect in the Pr-based heavy fermion superconductor PrOs$_4$Sb$_{12}$.
The topology of Fermi surface is close to the reference compound LaOs$_4$Sb$_{12}$ and well explained by the band structure calculation based on the FLAPW-LDA+U method, where the 4{\it f} electrons to be localized. We have confirmed a highly enhanced cyclotron effective mass 2.4$\sim$7.6$m_{\rm 0}$ which is apparently large compared to the usual Pr-based compounds.
\end{abstract}


\maketitle

The Pr-based compounds in the filled skutterudites have attracted much attention because of their exotic properties, such as metal-insulator transition at $T_{\rm MI}=60$~K in PrRu$_{4}$P$_{12}$\cite{Sekine} and unusual heavy fermion (HF) behavior in PrFe$_{4}$P$_{12}$.~\cite{Sato,Sugawara,Aoki} The quadrupolar interaction is thought to be a key mechanism to explain such anomalous behaviors. Especially for PrFe$_{4}$P$_{12}$, both the extraordinarily enhanced effective mass ($m^{\rm \ast}_{\rm c}=81m_{\rm 0}$) and the non-magnetic low field ordered state  below 6.5~K are inferred to originate from the quadrupolar interaction.~\cite{Sugawara,Aoki,Hao}

Recently, PrOs$_4$Sb$_{12}$ was reported to exhibit superconductivity below $T_{\rm C}=1.85~{\rm K}$.~\cite{Bauer,Maple} From the large specific heat jump at $T_{\rm C}$, $\Delta C/T_{\rm C}$$\sim$
500~mJ/K$^{2}\cdot$ mol, Bauer {\it et al.} claimed that PrOs$_4$Sb$_{12}$ is the first example of the Pr-based HF-superconductor.~\cite{Bauer,Maple} They also inferred that the quadrupolar interaction plays an important role in the HF-superconductivity, since the magnetic susceptibility, specific heat $C(T)$, and inelastic neutron scattering measurements suggest the crystal electric field (CEF) ground state to be a non-Kramers doublet carrying quadrupole moments. The recent Sb-NQR measurement on PrOs$_4$Sb$_{12}$ have revealed that the temperature $T$ dependence of nuclear-spin-lattice-relaxation rate $1/T_{1}$ shows neither a coherence peak nor $T^{3}$ like power-law dependence below $T_{\rm C}$.~\cite{Kotegawa} The muon-spin relaxation measurements suggest an isotropic superconducting (SC) energy gap,~\cite{MacLaughlin} while the anisotropy of thermal conductivity against the magnetic field directions indicates that the SC energy gap has point nodes.~\cite{Izawa} Such unusual properties cannot be consistently explained neither by a conventional isotropic nor any anisotropic gap superconductivity reported until now, suggesting some novel type of HF-superconductivity of this compound.~\cite{Bauer,Maple,Kotegawa,MacLaughlin,Izawa,Miyake}
In the normal state above $\sim4.5$~T, Aoki {\it et al}. reported an anomalous field-induced ordered phase (FIOP) based on the specific heat measurements; possibly a quadrupolar ordering.~\cite{Aoki_JPSJ}

In order to understand these unusual properties, it is essential to determine the electronic structure of this compound and to directly confirm the highly enhanced effective mass. 
In this paper, we report the de Haas-van Alphen (dHvA) experiment on PrOs$_4$Sb$_{12}$,~\cite{Sugawara_SCES02} which is the most powerful tool to clarify the Fermi surface (FS) topology and to determine the cyclotron effective mass. The experimental result is compared with the band structure calculation based on the FLAPW-LDA+U method.

Single crystals of PrOs$_4$Sb$_{12}$ and the reference LaOs$_4$Sb$_{12}$ were grown by Sb-self-flux method basically same as in ref.~\cite{Takeda,Bauer_CeOs4Sb12}, using high-purity elements, 4N (99.99\% pure)-Pr, 4N-La, 3N-Os and 6N-Sb.
The typical forms of the single crystals were cubic or rectangular shape with a largest dimension 
of about 3 mm.
The crystal structure of filled skutterudite, belonging to the space group ${\it Im}$\={3} ($T_h^5$, \#204), was verified by the powder neutron diffraction experiments performed by VEGA at KEK Tsukuba, Japan. The lattice constant of PrOs$_4$Sb$_{12}$ obtained from the Rietveld refinement is $a=9.30311$~\AA~, which is close to the reported value.~\cite{Braun} Fractional coordinates of Sb at the 24g site are determined as (0, 0.3405, 0.1561) which is used in the band structure calculation. 
The residual resistivity $\rho_{\rm 0}$ and the residual resistivity ratio (RRR) of the present samples are $\rho_{\rm 0}=8\mu\Omega\cdot$cm and ${\rm RRR}=55$ for PrOs$_4$Sb$_{12}$, and $\rho_{\rm 0}=2.8\mu\Omega\cdot$cm and ${\rm RRR}=100$ for LaOs$_4$Sb$_{12}$.
The dHvA experiments were performed in a top loading dilution refrigerator cooled down to 30 mK with a 17~T superconducting magnet. The dHvA signals were detected by means of the conventional field modulation method with a low frequency ($f\sim10$~Hz).

\begin{figure}[h]
\begin{center}\leavevmode
\includegraphics[width=0.8\linewidth]{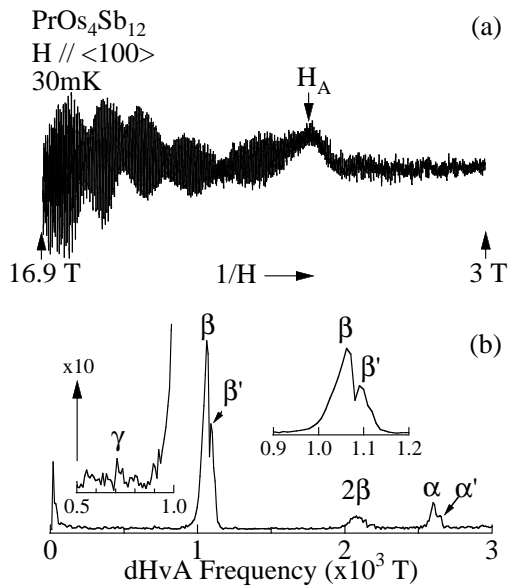}
\caption{ (a) Typical dHvA oscillations and (b) the 
FFT spectra in PrOs$_4$Sb$_{12}$.}
\label{Osc&FFT}
\end{center}
\end{figure}

Figure~\ref{Osc&FFT} shows (a) the typical dHvA oscillations and (b) the 
fast Fourier transformation (FFT) spectra in PrOs$_4$Sb$_{12}$ for the field $H$
along $\langle 100 \rangle$ direction.
The oscillations become detectable just above $H_{\rm C2}=2.2$~T and a faint peak anomaly appears at $H_{\rm A}\sim4.4$~T, indicating a slight slope change in the field dependence of magnetization $M(H)$.~\cite{Tenya} This field $H_{\rm A}$ almost agrees with the phase boundary of FIOP determined by the $C(T)$ measurement in the magnetic fields.~\cite{Aoki_JPSJ} Above  $H_{\rm A}$ a small spin-splitting is observed in the dHvA frequencies reflected as beats in the oscillations, which is frequently observed in the magnetic materials.~\cite{Onuki}
The three fundamental dHvA branches $\alpha$, $\beta$ and $\gamma$ are identified.

\begin{figure}[h]
\begin{center}\leavevmode
\includegraphics[width=0.8\linewidth]{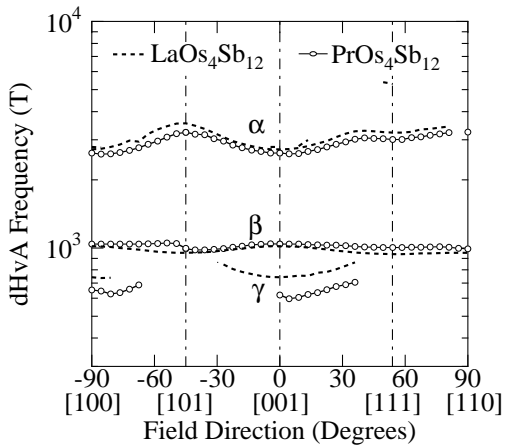}
\caption{Comparison of the angular dependence of the dHvA frequencies between PrOs$_4$Sb$_{12}$ (circles) and LaOs$_4$Sb$_{12}$ (dashed lines).}
\label{AngDepdHvA}
\end{center}
\end{figure}

Figure~\ref{AngDepdHvA} shows the angular dependence of dHvA frequency in PrOs$_4$Sb$_{12}$ along with that in LaOs$_4$Sb$_{12}$.
$\alpha$- and $\beta$-branches have been observed over the whole field directions in the cubic symmetry, indicating closed FSs. The branch $\gamma$ observed only in the limited angular ranges centered at $\langle 100 \rangle$ suggests a part of the multiply connected one. 
The striking resemblance of the angular dependence of dHvA frequencies between PrOs$_4$Sb$_{12}$ and LaOs$_4$Sb$_{12}$ indicates the closeness of FS topology between the two compounds, and evidences the well localized nature of 4$f$-electrons in PrOs$_{4}$Sb$_{12}$.

\begin{figure}[h]
\begin{center}\leavevmode
\includegraphics[width=0.8\linewidth]{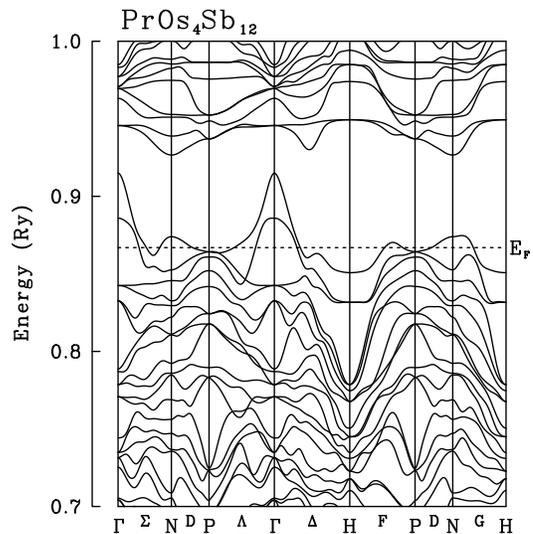}
\caption{Energy band structure in PrOs$_4$Sb$_{12}$.}
\label{Band}
\end{center}
\end{figure}

\begin{figure}[h]
\begin{center}\leavevmode
\includegraphics[width=0.8\linewidth]{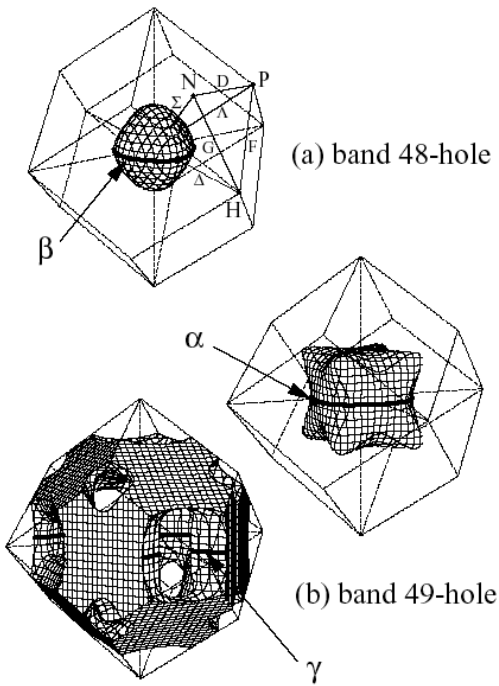}
\caption{Fermi surface of PrOs$_4$Sb$_{12}$.}
\label{FS}
\end{center}
\end{figure}

\begin{figure}[t]
\begin{center}\leavevmode
\includegraphics[width=1\linewidth]{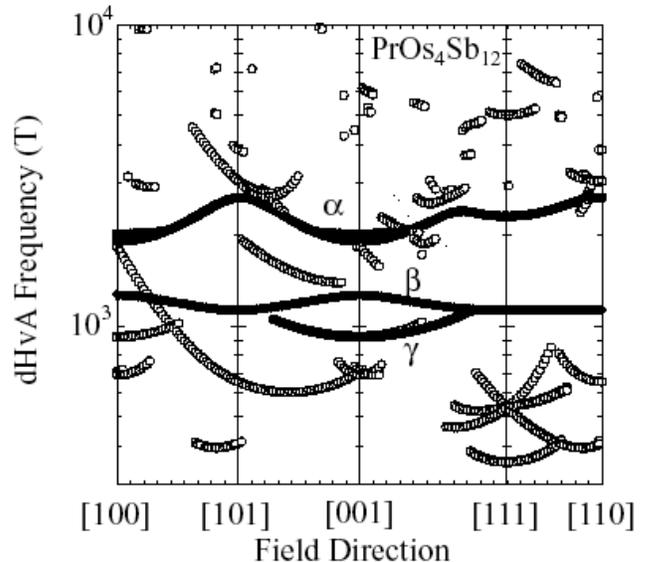}
\caption{Angular dependence of the theoretical dHvA branches in PrOs$_4$Sb$_{12}$. The dHvA branches indicated by the closed marks are identified by the experiments.}
\label{BanddHvA}
\end{center}
\end{figure}

In order to assign the origin of dHvA branches, the band structure calculation is carried out using an FLAPW and LDA+U method, in which the 4$f$-electrons in PrOs$_{4}$Sb$_{12}$ is treated as localized. Here we assume the $\Gamma_1$ singlet to treat the localized $4f^2$ electrons with the cubic symmetry.
The details of the calculation are described in ref.~\cite{Harima_JMMM_01,Harima_PhysB_02}.
Figures~\ref{Band} and \ref{FS} show the calculated energy band structure and the FS in PrOs$_4$Sb$_{12}$, respectively. 
The FS is composed of the 48th and 49th band hole sheets. The 48th band forms an almost spherical sheet which slightly projects along the $\langle 100 \rangle$ direction centered at the $\Gamma$ point. The 49th band gives a round cubic sheet centered at $\Gamma$ point and a multiply connected one whose main parts are centered at the N points.
Figure~\ref{BanddHvA} shows the angular dependence of the calculated dHvA
frequencies, which reasonably well explains all the observed dHvA branches. A slight disagreement in the absolute values is probably ascribed to the assumption of the $\Gamma_1$ singlet ground state and/or the effect beyond the LDA+U treatment.
Branch $\beta$ originates from the 48th-band spherical FS, branch $\alpha$ is due to the 49th-band round cube FS and branch $\gamma$ originates from the extremal orbit surrounding the N point of the part of multiply connected one. Many other dHvA branches predicted in the band structure calculation have not been observed in the present experiments. That can be understood as a large reduction of the dHvA amplitude due to the large curvature factors $A''$ of these branches; i.e., for $H\|\langle 100 \rangle$, $|A''|=0.08$ for the $\gamma$-branch whereas $|A''|=2.72$ for the dHvA branch exhibiting the frequency of $F=0.71\times10^3$~T.

As the most attractive information, we have estimated the cyclotron effective mass $m^{\rm \ast}_{\rm c}$ from the temperature dependence of the dHvA amplitude. The comparison of dHvA frequencies and $m^{\rm \ast}_{\rm c}$ between these compounds is given in Table~\ref{table1}.~\cite{Table1}
\begin{table*}
\caption{Comparison of the dHvA frequency $F$ and the cyclotron effective mass $m_{\rm c}^*$ between PrOs$_{4}$Sb$_{12}$ and LaOs$_{4}$Sb$_{12}$.}
\centering
\begin{tabular}{@{\hspace{\tabcolsep}\extracolsep{\fill}}cccccccc}
\hline
\multicolumn{2}{c}{} & \multicolumn{2}{c}{PrOs$_{4}$Sb$_{12}$(Exper.)} & \multicolumn{2}{c}{PrOs$_{4}$Sb$_{12}$(Theor.)} & \multicolumn{2}{c}{LaOs$_{4}$Sb$_{12}$} \\
Field direction & Branch & $F(\times10^3$~T) & $m_{\rm c}^*$($m_0$) & $F(\times10^3$~T) & $m_{\rm c}^*$($m_0$)    & $F(\times10^3$~T) & $m_{\rm c}^*$($m_0$)  \\ 
\hline 
     $H\|\langle 100 \rangle$ & $\alpha$  & 2.61  & 4.1 & 1.91  & 0.88  & 2.79  & 2.5 \\		
     & $\beta$  &1.07  & 2.5 & 1.27  & 0.45 & 1.02  & 0.71 \\
     & $\gamma$ & 0.71  & 7.6 & 0.93  & 0.79 & 0.74  & 2.8 \\
\hline 
     $H\|\langle 110 \rangle$ & $\alpha$ & 3.24  & 4.9 & 2.68  & 1.82 & 3.55  & 4.1 \\		
     & $\beta$  & 0.89  & 3.9 & 1.13  & 0.39 & 0.95  & 0.65\\
\hline 
     $H\|\langle 111 \rangle$ & $\alpha$ & 3.00  & 5.8 & 2.32 & 1.27 & 3.23 & 2.8 \\		
     & $\beta$ & 0.86  & 2.4 & 1.13  & 0.38  & 0.94  & 0.65 \\ 
\hline
\end{tabular}
\label{table1}
\end{table*}
In contrast with the closeness of the FS topology, the cyclotron effective masses are up to $\sim6$ times enhanced compared to those in LaOs$_{4}$Sb$_{12}$. 
The mass enhancement is apparently larger than those in the ordinary Pr-based compounds such as PrIn$_3$ and PrSb,~\cite{Onuki} although it is not so large as that in PrFe$_{4}$P$_{12}$.~\cite{Sugawara}
For the comparison with the mass enhancement estimated from the specific heat measurement, 
we simply estimate the Sommerfeld coefficient from the FS volume and $m^{\rm \ast}_{\rm c}$ in the present experiments assuming spherical FSs. The estimated value of $\sim 150$mJ/K$^{2}\cdot$mol is still a factor of 2 or more smaller than that in the specific heat measurement; Bauer {\it et al.} estimated the Sommerfeld coefficient of PrOs$_{4}$Sb$_{12}$ to be $350\sim 750$~mJ/K$^{2}\cdot$mol.~\cite{Bauer}

We have no definitive explanation for this discrepancy only from the present information. Nevertheless, a possible origin may be hidden in the difference of the measuring conditions, such as magnetic field and temperature; the Sommerfeld coefficient in the specific heat was estimated at 0~T, and around $T_{\rm C}$ close to the Schottky-like peak at 3~K, while the dHvA measurements were made at high fields of $3\sim17$~T and lower temperatures of $30\sim400$~mK. If the peak is associated with the nearby magnetic CEF excited state situated about 8~K above the ground state,~\cite{Bauer,Aoki_JPSJ} an additional enhancement of effective mass could be expected from the magnetic instability. As another origin, the spin-splitting effect should not be discarded. In the present experiments we have estimated $m^{\rm \ast}_{\rm c}$ only for the larger amplitude spin-direction, since the insufficient experimental resolution prevents to determine the masses for the higher frequencies ($\alpha'$ and $\beta'$ as shown in Fig.~\ref{Osc&FFT}(b)). Only for $\beta$-branch with the large amplitude, we could roughly estimate $m^{\rm \ast}_{\rm c}$ of the higher frequency branch $\beta'$, which is about 20\%  enhanced compared with that of the lower one $\beta$. The larger mass enhancement for the higher frequency branch might explain the discrepancy of the mass enhancement between the Sommerfeld coefficient and $m^{\rm \ast}_{\rm c}$.

It might be informative to compare PrOs$_4$Sb$_{12}$ with another Pr-based filled skutterudite superconductor PrRu$_4$Sb$_{12}$ as shown in Table~\ref{table2}.~\cite{Sugawara_ROs4Sb12}
\begin{table*}
\caption{Comparison of the superconducting critical temperature $T_{\rm C}$, superconducting specific heat jump $\Delta C$ divided by $T_{\rm C}$ ($\Delta C/T_{\rm C}$), Sommerfeld coefficient, and effective mass $m_{\rm c}^*$ in RT$_{4}$Sb$_{12}$ (R=La, Pr, T=Ru, Os).}
\centering
\begin{tabular}{@{\hspace{\tabcolsep}\extracolsep{\fill}}lcccc}
\hline 
\multicolumn{1}{c}{} & \multicolumn{1}{c}{PrOs$_{4}$Sb$_{12}$} & \multicolumn{1}{c}{LaOs$_{4}$Sb$_{12}$} & \multicolumn{1}{c}{PrRu$_{4}$Sb$_{12}$} & \multicolumn{1}{c}{LaRu$_{4}$Sb$_{12}$} \\
\hline 
     $T_{\rm C}$ (K) & 1.85~\cite{Bauer,Maple} & 0.74~\cite{Kotegawa,Sugawara_ROs4Sb12} & 1.04~\cite{Takeda} & 3.58~\cite{Takeda} \\
     $\Delta C/T_{\rm C}$ (mJ/K$^{2}\cdot$mol)  & 500~\cite{Bauer,Maple} & 84~\cite{Sugawara_ROs4Sb12} & 110~\cite{Takeda}  & 82~\cite{Takeda} \\
     Sommerfeld coefficient (mJ/K$^{2}\cdot$mol) & $350\sim750$~\cite{Bauer,Maple} & 36,~\cite{Bauer_CeOs4Sb12} 56~\cite{Sugawara_ROs4Sb12} & 59~\cite{Takeda} & 37~\cite{Takeda} \\
 $m_{\rm c}^*/m_0$ for $\gamma$-branch & 7.6  & 2.8 & 1.6~\cite{Matsuda} & 1.4~\cite{Matsuda} \\
\hline
\end{tabular}
\label{table2}
\end{table*}
The localized character of 4$f$-electrons, namely the closeness of the FSs with those in LaRu$_4$Sb$_{12}$, has been confirmed also in PrRu$_4$Sb$_{12}$ based on the dHvA experiment.~\cite{Matsuda} On the contrary, the mass enhancement is quite small in PrRu$_4$Sb$_{12}$ which is in sharp contrast to that in PrOs$_4$Sb$_{12}$. 
For PrOs$_4$Sb$_{12}$, the CEF ground state was inferred to be a non-Kramers doublet carrying quadrupole moments.~\cite{Bauer,Maple} On the other hand, Takeda and Ishikawa inferred PrRu$_4$Sb$_{12}$ to have the $\Gamma_{\rm 1}$ singlet ground state.~\cite{Takeda_JPSJ} On the $T_{\rm C}$ compared with the La-references, the two compounds have different sense; $T_{\rm C}$ for PrOs$_4$Sb$_{12}$ is higher than that for La-reference, that is unusual if we take into account that PrOs$_4$Sb$_{12}$ contains the magnetic element Pr.
It should be also noted that the superconductivity observed in PrRu$_4$Sb$_{12}$ is the ordinary BCS-type.~\cite{Takeda_JPSJ,Kotegawa} 
These facts naturally indicate the essential role of the mass enhancement associated with 4$f$-electrons in the superconductivity of PrOs$_4$Sb$_{12}$.

In summary, we have directly confirmed the heavy mass in PrOs$_4$Sb$_{12}$ by the dHvA experiment, indicating PrOs$_4$Sb$_{12}$ to be the first member of HF-superconductor as Pr-compound. However, the well localized character of 4$f$-electrons, evidenced in the closeness of the FS to those in LaOs$_{4}$Sb$_{12}$, indicates the uniqueness of this compounds compared to the typical HF-superconductors such as UPt$_3$ and CeCoIn$_5$,~\cite{Kimura,Settai} where the itinerant $f$-electron model is applicable. From the comparison with another Pr-based filled skutterudite superconductor PrRu$_4$Sb$_{12}$, the quadrupolar interaction associated with the non-Kramers doublet CEF ground state is thought to be responsible for the HF-superconductivity in PrOs$_4$Sb$_{12}$.

We thank Dr.~H.~Kotegawa, Professor~Y.~Kitaoka, Professor~M.~B.~Maple, and Professor~K.~Miyake for the helpful discussion. This work was supported by the Grant-in-Aid for Scientific Research from the Ministry of Education, Science, Sports and Culture of Japan.


\begin{thebibliography}{}
\bibitem{Table1}The dHvA frequencies and $m^{\rm \ast}_{\rm c}$ plotted in Fig.~\ref{AngDepdHvA} and listed in Table~\ref{table1} are average values over the field ranges $10-12$~T for PrOs$_4$Sb$_{12}$ and $8-12$~T for LaOs$_4$Sb$_{12}$.
\end{thebibliography}
\end{document}